# Deep learning-based attenuation correction in the image domain for myocardial perfusion SPECT imaging

Samaneh Mostafapour[1], Faeze Gholamiankhah[2], Sirvan Maroofpour[3], Mahdi Momennezhad[4], Mohsen Asadinezhad[1], Seyed Rasoul Zakavi[4], Hossein Arabi[5]

[1]Department of Radiology Technology, Faculty of Paramedical Sciences, Mashhad University of Medical Sciences, Mashhad, Iran

[2]Department of Medical Physics, Faculty of Medicine, Shahid Sadoughi University of Medical Sciences, Yazd, Iran

[3]Department of Medical Physics, Faculty of Medicine, Mashhad University of medical sciences, Mashhad, Iran

[4]Nuclear Medicine Research Center, Mashhad University of Medical Sciences, Mashhad, Iran

[5]Division of Nuclear Medicine and Molecular Imaging, Department of Medical Imaging, Geneva University Hospital, CH-1211 Geneva 4, Switzerland

**Short title:** Deep learning-based attenuation correction of MPI-SPECT images

**Abstract**

**Objective:**

In this work, we set out to investigate the accuracy of direct attenuation correction (AC) in the image domain for the myocardial perfusion SPECT imaging (MPI-SPECT) using two residual (ResNet) and UNet deep convolutional neural networks.

**Methods:**

The MPI-SPECT $^{99m}$Tc-sestamibi images of 99 participants were retrospectively examined. UNet and ResNet networks were trained using SPECT non-attenuation corrected images as input and CT-based attenuation corrected SPECT images (CT-AC) as reference. The Chang AC approach, considering a uniform attenuation coefficient within the body contour, was also implemented to provide a baseline to assess the performance of the deep learning models. Quantitative and clinical evaluation of the proposed methods were performed considering SPECT CT-AC images of 19 subjects (external validation set) as reference using the mean absolute error (MAE), relative error (RE), structural similarity index, peak signal-to-noise ratio (PSNR) metrics, as well as relevant clinical indices such as perfusion deficit (TPD).

**Results:**

Overall, the deep learning solution exhibited good agreement with the CT-based AC, noticeably outperforming the Chang method. The ResNet and UNet models resulted in the ME (count) of -6.99±16.72 and -4.41±11.8 and SSIM of 0.99±0.04 and 0.98±0.05, respectively. While the Change approach led to ME and SSIM of 25.52±33.98 and 0.93±0.09, respectively. Similarly, the clinical evaluation revealed a mean TPD of 12.78±9.22 and 12.57±8.93 for the ResNet and UNet models, respectively, compared to 12.84±8.63 obtained from the reference SPECT CT-AC images. On the other hand, the Chang approach led to a mean TPD of 16.68±11.24.

**Conclusion:**

We evaluated two deep convolutional neural networks to estimate SPECT-AC images directly from the non-attenuation corrected images. The deep learning solutions exhibited the promising potential to generate reliable attenuation corrected SPECT images without the use of transmission scanning.

# 1. INTRODUCTION

Single-photon emission computed tomography (SPECT) myocardial perfusion imaging (MPI) is a widely used diagnostic modality in cardiovascular clinical diagnosis. This method as a non-invasive imaging examination plays an important role in the evaluation of myocardial ischemia, coronary artery disease (CAD), and its risk classification. SPECT imaging reveals cardiac physiology which would lead to earlier detection of pathophysiology or cardiac damages before the occurrence of the morphological damages, wherein it is more likely the cardiac damage is reversed (particularly in the early stages) [1-3].

The quality of the images obtained from SPECT-MPI plays a critical role in the diagnosis procedure, greatly influencing the diagnostic or prognostic accuracy. In nuclear medicine imaging in general and SPECT imaging in particular, there are a number of degradation factors that limit the quality and the quantitative accuracy of the reconstructed images such as photons attenuation, Compton scatter, the partial-volume effect as a result of limited spatial resolution, and septal penetration. In the last decades, the compensation of the above-mentioned degrading factors has been an ongoing challenge aiming at improving the overall diagnostic accuracy and efficiency of the SPECT systems [4-8].

Attenuated and scattered photons within the body would greatly affect the accuracy and specificity of myocardial perfusion imaging since they vary noticeably (in terms of distribution and intensity) from one patient (scan) to another (scan). It also diminishes the accuracy of the quantification and the physicians' confidence for interpretation of images via causing pseudo uptakes or defects. This effect is originated from the attenuation of photons within the human body that varies significantly in inhomogeneous tissues such as the thorax and causes an apparent decrease in the activity of the reconstructed images. The diaphragm in men, breast in women, lateral chest walls, and abdomen in patients with large body mass index can be the most common source of photon attenuation artifacts in myocardial perfusion SPECT imaging. The photon attenuation is more significant at the low energy photons which reduces the true value of activity concentration and may present with a pseudo perfusion or metabolism defect [5, 9, 10].

So far, several techniques and methods have been suggested/employed to diminish the adverse impact of the photon attenuation in myocardial perfusion imaging, such as prone imaging [11, 12], electrocardiography (ECG) gated SPECT imaging [11], and compensation of the attenuated photons within image reconstruction [13]. The gated MPI provides additional functional information which distinguishes between myocardial infarctions and fixed myocardial perfusion defects due to soft-tissue attenuation; however, this technique has limitations in patients with subendocardial myocardial infarctions [11]. Prone imaging is helpful in diaphragmatic attenuation, but it faces the limitations such as requiring additional acquisition, loss of image contrast, and inefficiency in dealing with breast attenuation [5, 12]. Performing attenuation correction (AC) in SPECT-MPI improves diagnostic accuracy and normalcy rate which helps physicians to interpret the images confidently; and also eliminates the need for a rest study in normal patients [5, 14, 15].

Attenuation correction in SPECT imaging has been performed using mainly two generic approaches; transmission-less and transmission-based methods. In transmission-based methods, an external radionuclide source or Computed Tomography (CT) scan is used to generate a patient-specific attenuation map which is regarded as the standard/reference AC approach [16]. Since the advent of the hybrid SPECT and CT scanners, combining these two modalities into a single imaging system, CT-based AC has become the commonly used and standard AC approach in SPECT imaging. This method besides providing high-resolution structural images (complementary anatomical information to functional imaging) is able to generate patient-specific AC maps with relatively low noise and high image quality [10]. However, CT-based AC maps may suffer from misalignment errors between emission and transmission scans due to involuntary or voluntary patient movements as well as increased patients' radiation dose [17-19].

Transmission-less approaches generate attenuation maps through defining body contour and assuming a uniform distribution of attenuation coefficients within the body. Attenuation correction factors could also be estimated from the measured emission data. These approaches suffer from a lack of patient specificity (ignoring the inhomogeneity of tissues within the body) or high noise levels [13, 19]. Given the structural magnetic resonance (MR) images obtained from the hybrid SPECT-MRI scanners, synthetic CT images could be estimated from the MR images to perform AC within SPECT image reconstruction [20].

The same challenge is faced in Positron Emission Tomography (PET) imaging, wherein serious attempts have been made to cope with this issue in the absence of the standard transmission or CT scan [21, 22]. These methods, mostly developed for hybrid PET/MR scanners, rely on the accompanying structural MR images to generate synthetic CT images for the task of PET attenuation and scatter correction. These include the segmentation-based [23, 24], atlas-based [25, 26], joint attenuation and emission reconstruction [27, 28] in PET/MR as well as standalone PET scanners. Recently, artificial intelligence-based methods and in particular deep learning algorithms have revolutionized the generation of synthetic images [29, 30]. Deep learning approaches enabled the direct generation of accurate synthetic CT images from MR images [31, 32], estimation of the patient-specific AC map from the emission data [33], as well as the application of the direct attenuation and scatter correction in the images domain without the need for any anatomical images [34, 35].

In this work, we set out to investigate the accuracy of direct attenuation correction in the image domain for the SPECT myocardial perfusion imaging using the residual and UNet deep neural networks. These deep learning models were trained to directly estimate attenuation corrected SPECT images from non-attenuation corrected ones without the use of any anatomical images. We also implemented the Chang method as a baseline for comparison of the different attenuation correction approaches of the MPI-SPECT images. The Chang method, used in SPECT-alone systems, assumes a uniform attenuation coefficient for the entire area within the body contour [36-38]. CT-based attenuation correction was considered as the standard/reference method to quantitatively and clinically evaluate the deep learning-based and Chang AC approaches.

## 2. MATERIALS AND METHODS

### 2.1 SPECT/CT data acquisition

This retrospective study employed 99 clinical subjects including both normal and abnormal patients. The study was approved by the ethical committee of Mashhad University of Medical Sciences (ethic number IR.MUMS.REC.1398.235). The entire patients were scanned at the same nuclear medicine center with $^{99m}$Tc-sestamibi for Gated MPI-SPECT stress studies. The patients were injected with 740-925 MBq activity of $^{99m}$Tc-sestamibi and acquisition was performed 45-60 min after the intravenous injection. SPECT and CT images were acquired on a Discovery NM/CT 670 (GE Healthcare) SPECT/CT scanner. CT images were acquired with the tube potential of 120 kVp, tube current of 50 mAs, slice thickness of 5 mm, matrix size of 512 × 512, and a voxel size of 0.97 × 0.97 × 5 mm$^3$. The reconstruction of both attenuation and non-attenuation corrected SPECT images were performed using the ordered subsets expectation maximization (OSEM) algorithm with 4 iterations, 10 subsets, and a post-reconstruction Butterworth filter (frequency of 0.4 and power of 10) and the size of the SPECT images was 64 × 64 × 19. Electrocardiogram (ECG)-gated MPI-SPECT data were acquired using 8 frames per cardiac cycle for R–R interval length using forward-backward gating method. The SPECT data were acquired with a low-energy high-resolution (LEHR) collimator using 30 projections with 25 seconds per projection covering a 180-degree rotation.

### 2.2 Network architecture

In this work, two deep learning models, ResNet and UNet, were implemented in Tensorflow using the NiftyNet platform to predict SPECT CT-AC from SPECT non-AC images in the image domain. The NiftyNet infrastructure is an open-source platform, developed upon the TensorFlow module in the Python environment, that provides a high-level deep learning pathway for medical image analysis and processing, including segmentation, classification, and image regression [39].

The ResNet architecture, illustrated in Figure 1 [40], consists of 20 convolutional layers that all layers except the last one, which is a fully connected softmax layer, have 3×3×3 kernels, batch normalization, rectified linear unit (ReLU), and a shortcut connection added to each pair of 3×3×3 filters to enhance the training speed and efficiency. The structure benefits from dilated convolutions (by a factor of two for the middle seven layers and a factor of four for the last six layers (Figure 1)) that support the exponential inflation of the receptive field while preserving the original resolution of the input images.

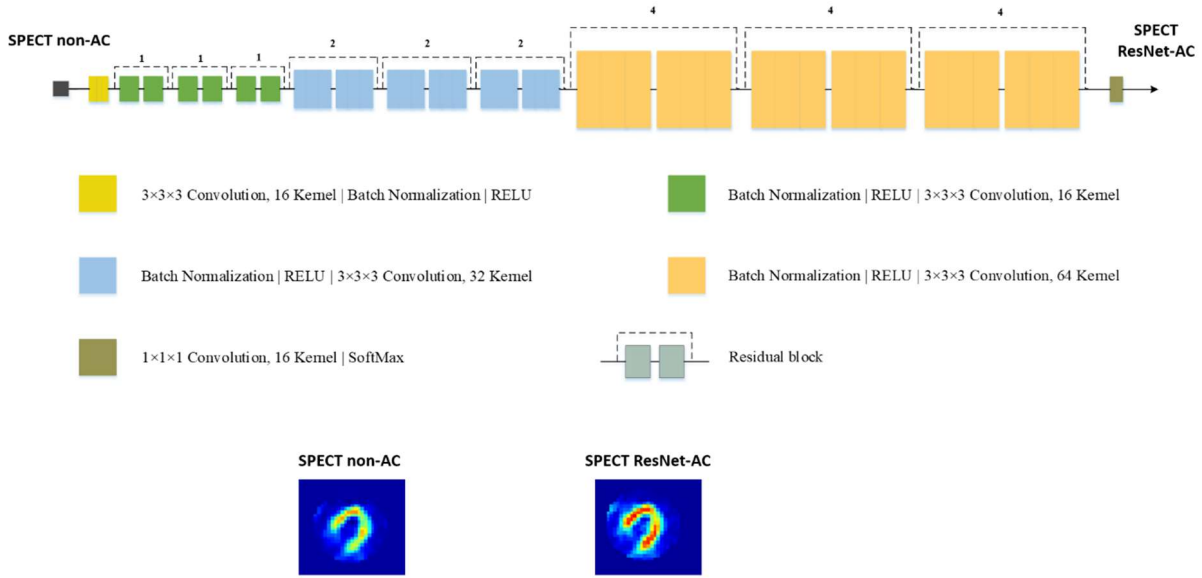

**Figure 1.** Architecture of ResNet network implemented in the NiftyNet platform.

The UNet architecture, implemented in the NiftyNet platform, [41], is an upgraded version of the traditional convolutional neural network with a 'U' shape and symmetric structure that consists of 23 convolutional layers in two major parts. The first part includes the contracting path (left part in Figure 2) which follows the general convolutional process, each consists of two 3×3 convolution kernels followed by a rectified linear unit (ReLU) and a 2×2 max-pooling operation. At each downsampling step (max-pooling operation) the number of channels increases as the convolution process will increase the depth of the image. The second part is the expansive path (right part in Figure 2) which is consists of an upsampling of the feature maps (reducing the number of channels at every step) followed by a transposed 2×2 convolutional layer (concatenated with the corresponding feature map from the contracting path), two 3×3 convolutions, and a ReLU.

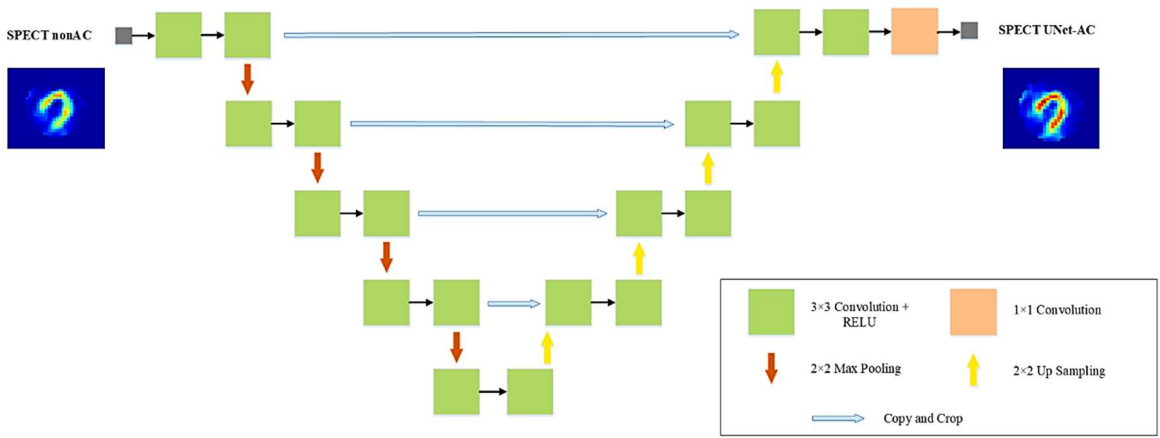

**Figure 2.** Architecture of UNet network implemented in the NiftyNet platform.

## 2.3 Implementation details

The training of the networks for the prediction of SPECT CT-AC images was performed using 80 pairs of SPECT non-AC and SPECT CT-AC images as input and output, respectively, and 19 SPECT non-AC images were used for the external validation dataset (this dataset was never used in the training). 5% of the training dataset was dedicated to evaluating the model within the training to verify and avoid the risk of overfitting. SPECT images' pixel values were normalized to 90% of their corresponding cumulative histograms to reduce the variation of intensity range across different patients. Furthermore, irrelevant voxels containing background air were removed through cropping the images into a matrix size of 64×64 voxels for the training of ResNet and UNet models to reduce the computational burden.

The ResNet model was trained using the following parameters; learning rate = 0.0001, optimizer = Adam, loss function = L2 Loss, decay = 0.0001, batch size = 40 and total iteration of 19 k. The training for the UNet model was performed using the same loss function, decay, and batch size, but the learning rate was 0.001. The training of the UNet was completed after almost 18 K iterations since its training loss reached its plateau. It should be noted that the training of these models was performed in 2-dimensional (2D) mode (using 2D slices).

## 2.4 Evaluation strategy

Quantitative evaluation of the synthetic SPECT images generated by the ResNet and UNet models and the SPECT Chang-AC images were performed against the reference SPECT CT-AC images. The Chang AC method was also implemented to provide a baseline for a comprehensive performance assessment of the deep learning models. The Chang method assumes a uniform attenuation coefficient of 0.09 cm$^{-1}$ (considering the thorax area with the lung tissue bearing low photon attenuation factors [42]) within the body contour. An empirical intensity threshold was applied to define the correct body contours (visually verified).

The quality of MPI SPECT-AC images was estimated using the following quantitative metrics; voxel-wise mean error (ME) (Eq. 1), mean absolute error (MAE) (Eq. 2), root mean square error (RMSE) (Eq. 3), and relative error (RE%) (Eq. 4) calculated between SPECT CT-AC and SPECT-AC images (generated by the two deep learning models and the Chang method).

$$ME = \frac{1}{N}\sum_{i=1}^{N} dS(i) \tag{1}$$

$$MAE = \frac{1}{N}\sum_{i=1}^{N} |dS(i)| \tag{2}$$

$$RMSE = \sqrt{\frac{1}{N} \sum_{i=1}^{N} (dS(i))^2} \tag{3}$$

$$RE(\%) = \frac{1}{N} \sum_{i=1}^{N} \frac{(S_C(i) - S_R(i))}{S_R(i)} \times 100 \tag{4}$$

Here, $N$ indicates the total number of voxels in the volume of interest, and by assuming $dS(i) = (S_C(i) - S_R(i))$, $S_R(i)$ is the reference image (SPECT CT-AC), $S_C(i)$ is the attenuation corrected images by either of the three different methods, $i$ stands for the voxel index in the SPECT-AC and SPECT CT-AC images. Moreover, the peak signal-to-noise ratio (PSNR) and structural similarity index (SSIM) were calculated between the different SPECT-AC images versus the reference SPCET CT-AC images using equations (5) and (6), respectively.

$$PSNR = 10 \log \left( \frac{I^2}{MSE} \right) \tag{5}$$

$$SSIM = \frac{(2\mu_R \mu_C + K_1)(2\delta_{R,C} + K_2)}{(\mu_R^2 + \mu_C^2 + K_1)(\delta_R^2 + \delta_C^2 + K_2)} \tag{6}$$

In Eq. (5), $I$ represents the maximum intensity value of either SPECT CT-AC or different SPECT-AC, whereas *MSE* is the mean square error. In Eq. (6), $\mu_R$ and $\mu_C$ indicate the mean value of SPECT CT-AC and SPECT-AC images, respectively. $\delta_R$ and $\delta_C$ are the variances of SPECT CT-AC and SPECT-AC images, whereas $\delta_{R,C}$ indicates their covariance. The parameters $K_1 = (k_1 I)^2$ and $K_2 = (k_2 I)^2$ with constants $k_1 = 0.01$ and $k_2 = 0.02$ were introduced to avoid division by very small numbers. Furthermore, the voxel-wise correlation of the tracer uptake between the ResNet, UNet, and Chang methods versus the reference SPECT CT-AC images was estimated using joint histogram analysis.

Clinical evaluation was performed based on the quantitative parameters extracted from Cedars-Sinai Medical Center software, Quantifed Perfusion SPECT (QPS). These quantitative parameters include Defect, Extent, Summed Stress Percent (SS%), Summed Stress Score (SSS), Total Perfusion Deficit (TPD%), Volume, Area, Shape Eccentricity (ECC), and Shape Index (SI).

## 3. RESULTS

Figure 3 provides sections along the short and long axes for a representative clinical subject, comparing SPECT ResNet-AC, SPECT UNet-AC, SPECT Chang-AC along with the corresponding SPECT CT-AC. Figure 3 shows that predicted images of deep learning models bear image quality comparable to the reference SPECT CT-AC, while noticeable differences are observed in Chang-AC images.

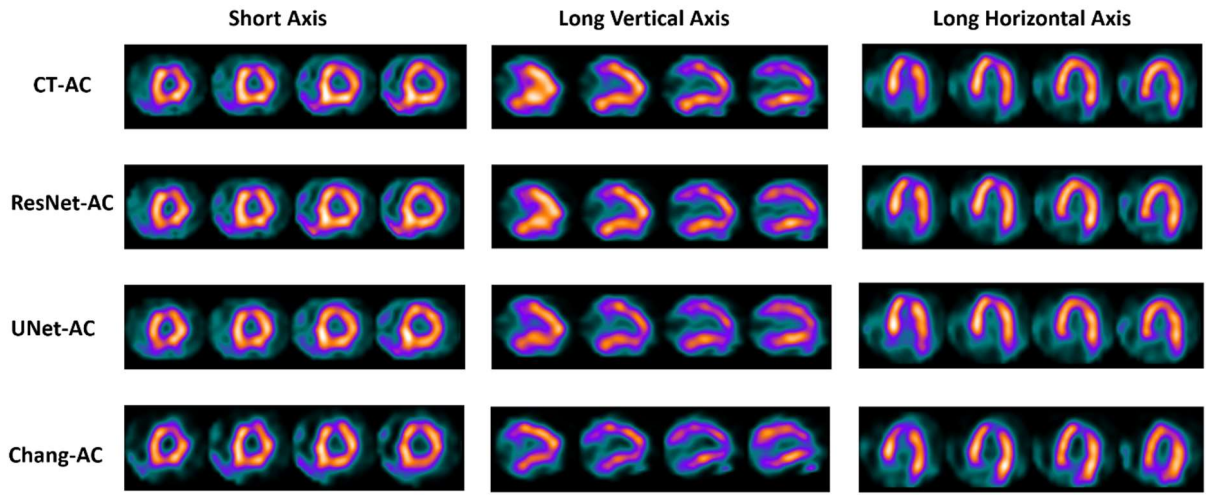

**Figure 3.** Representative MPI-SPECT images showing SPECT CT-AC images, attenuation corrected SPECT images generated by the ResNet and UNet models, and attenuation corrected SPECT images using Chang method (from top to bottom, respectively).

Representative long vertical axis views of the SPECT CT-AC, SPECT ResNet-AC, SPECT UNet-AC, and SPECT Chang-AC images along with the relative bias (%) maps with respect to the reference SPECT CT-AC images are shown in Figure 4. SPECT ResNet-AC and SPECT UNet-AC revealed good agreement in terms of signal recovery and underlying uptake pattern; while, the SPECT Chang-AC images exhibited a remarkable underestimation/overestimation of the activity concentration considering the horizontal profile drawn through the myocardium on the SPECT CT-AC, SPECT ResNet-AC, SPECT UNet-AC, and SPECT Chang-AC images (Figure 4).

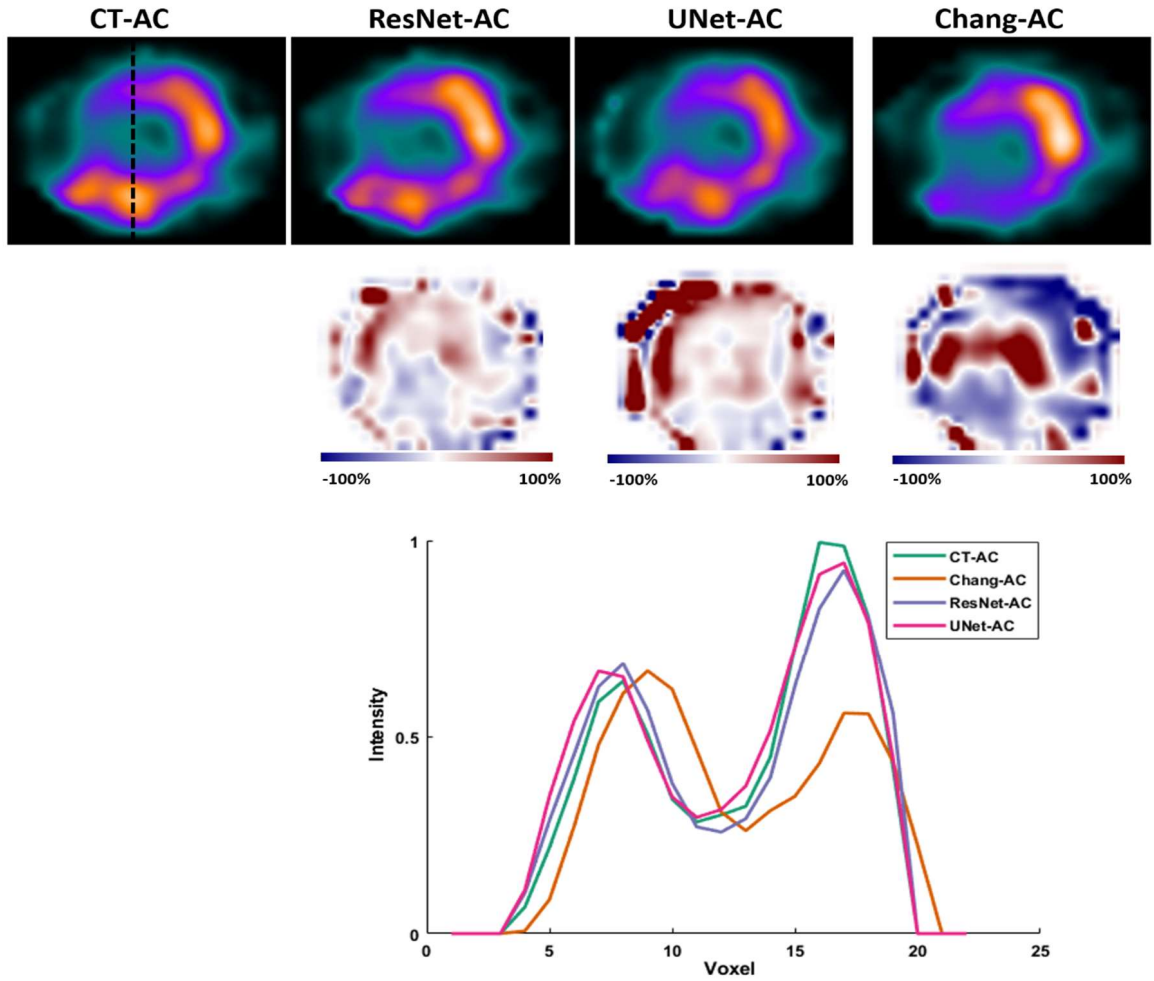

**Figure 4.** Representative long vertical axis views of SPECT CT-AC, SPECT ResNet-AC, SPECT UNet-AC, and SPECT Chang-AC along with relative bias (%) maps with respect to the reference SPECT CT-AC images. Horizontal line profiles drawn through the myocardium are also shown.

Table 1 summarizes the results of quantitative metrics (mean ± SD), including ME, MAE, RMSE, RE, SSIM, and PSNR, calculated on the SPECT MPI images attenuation corrected using the different methods, across the 19 patients of the external validation dataset. The SPECT UNet-AC revealed the smallest voxel-wise ME=-4.41±11.85, MAE=13.65±11.23, and RMSE=42.33±32.41, while ResNet-AC and Chang-AC methods led to ME of -6.99±16.72 and 25.52±33.98, MAE of 20.24±17.63 and 80.27±47.56, and RMSE of 42.33±32.41 and 83.06±27.98, respectively. The RE (%) metric revealed an underestimation of -0.34±15.03% in the SPECT ResNet-AC, and an overestimation of 0.42±15.88%, 11.74±31.67 in SPECT UNet-AC and SPECT Chang-AC images, respectively. The SPECT ResNet-AC and SPECT UNet-AC resulted in SSIM of 0.99±0.04, 0.98±0.05, and PSNR of 28.15±4.17 and 29.44±3.45, respectively.

**Table 1.** Statistical analysis of the image quality metrics measured in the attenuation corrected MPI-SPECT images by the different methods with respect to the reference SPECT CT-AC images.

| Methods | ME (counts) | MAE (counts) | RMSE (counts) | RE(%) | SSIM | PSNR |
|---|---|---|---|---|---|---|
| ResNet-AC | -6.99±16.72 | 20.24±17.63 | 42.33±32.41 | -0.34±15.03 | 0.99±0.04 | 28.15±4.17 |
| UNet-AC | -4.41±11.85 | 13.65±11.23 | 36.10±26.1 | 0.42±15.88 | 0.98±0.05 | 29.44±3.45 |
| Chang-AC | 25.52±33.98 | 80.27±47.56 | 83.06±27.98 | 11.74±31.67 | 0.93±0.09 | 22.06±2.50 |

Figure 5 presents the box plots of ME, MAE, RMSE, RE (%), RMSE, and SSIM metrics calculated for the different MPI-SPECT images.

The joint histogram analysis illustrating the voxel-wise correlation between the reference SPECT CT-AC images and attenuation corrected SPECT images using the different methods (Figure 6), demonstrates that SPECT images attenuation corrected by the UNet and ResNet models are highly correlated with the reference CT-based AC ($R^2$=0.99).

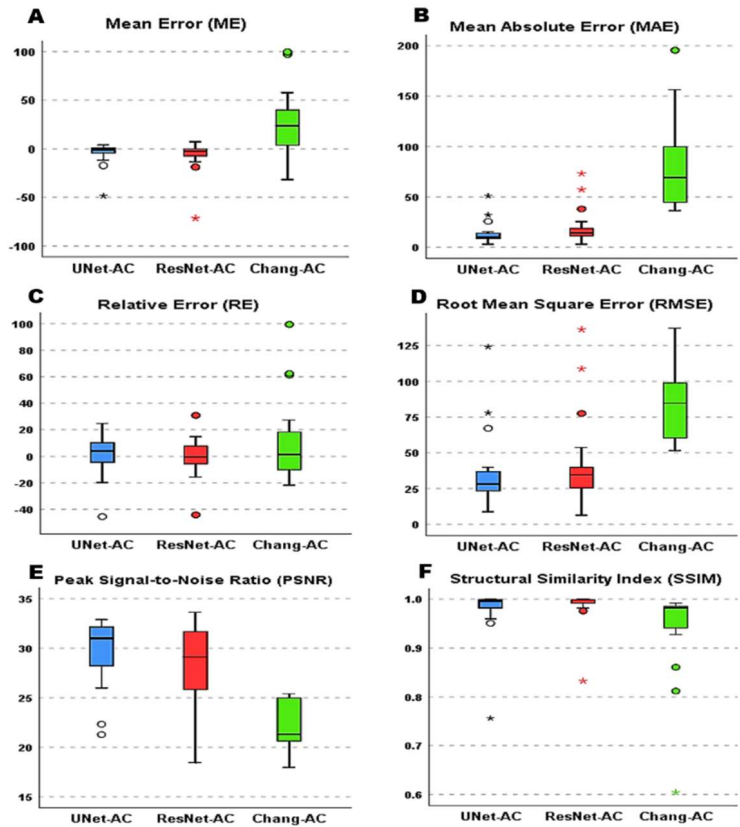

**Figure 5**. Comparison of (A) mean error (ME), (B) mean absolute errors (MAE), (C) relative errors (RE%), (D), root-mean-square errors (RMSE), (E) peak signal to noise ratio (PSNR), and (F) structural similarity index (SSIM), measured in the attenuation corrected SPECT images for the different methods (versus the reference SPECT CT-AC).

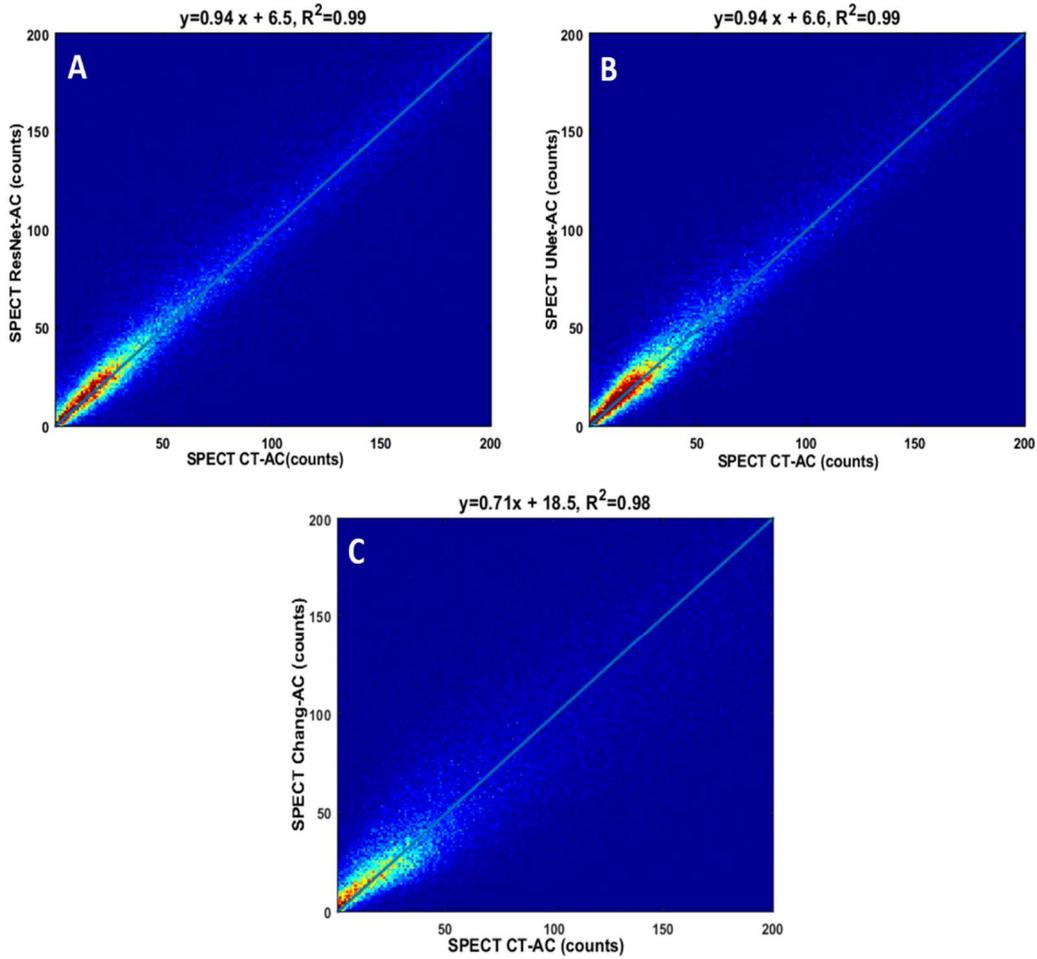

**Figure 6.** Voxel-wise joint correlation histogram analysis of attenuation corrected SPECT images by the different methods versus the reference SPECT CT-AC image; (A) ResNet, (B) UNet, and (C) Chang methods.

Tables 2 summarize the results of the statistical analysis obtained from the quantitative assessment of perfusion MPI-SPECT images. Regarding the TPD (%) index, SPECT images reconstructed from the ResNet (12.78±9.22) and UNet (12.57±8.93) models exhibited good agreement with the reference CT-AC images (TPD = 12.84±8.63). Moreover, detailed information about the distribution of TPD (%), SS (%), and SSS indices are shown by the box plots in Figure 7.

**Table 2.** Descriptive statistics of quantitative perfusion analysis of the MPI-SPECT images for the different attenuation corrected images using Cedars-Sinai-QPS software.

| Methods | | SSS | SS (%) | TPD (%) | Defect (cm²) | Extend (%) | Volume (ml) | Area (cm²) | SI | ECC |
|---|---|---|---|---|---|---|---|---|---|---|
| CT-AC | Mean ± SD | 9.94±6.31 | 14.31±9.04 | 12.84±8.63 | 21.57±17.24 | 15.84±11.07 | 89.89±41.52 | 130.63±32.66 | 0.59±0.09 | 0.82±0.05 |
| | Min | 0 | 0 | 2 | 2 | 1 | 26 | 75 | 0.43 | 0.72 |
| | Max | 25 | 37 | 35 | 56 | 42 | 185 | 191 | 0.75 | 0.91 |
| ResNet-AC | Mean ± SD | 9.68±6.12 | 14.21±9.07 | 12.78±9.22 | 20.78±17.52 | 15.73±11.36 | 87.57±39.72 | 127.94±31.82 | 0.62±0.10 | 0.81±0.04 |
| | Min | 1 | 1 | 1 | 1 | 1 | 24 | 72 | 0.44 | 0.75 |
| | Max | 25 | 37 | 36 | 56 | 42 | 178 | 183 | 0.80 | 0.90 |
| UNet-AC | Mean ± SD | 8.89±6.50 | 13±9.63 | 12.57±8.93 | 21.10±19.33 | 14.89±12.10 | 92.05±42.05 | 132.47±32.33 | 0.60±0.09 | 0.83±0.03 |
| | Min | 3 | 4 | 3 | 2 | 2 | 23 | 70 | 0.42 | 0.75 |
| | Max | 26 | 38 | 34 | 61 | 42 | 191 | 197 | 0.80 | 0.90 |
| Chang-AC | Mean ± SD | 12.94±9.15 | 19.05±13.37 | 16.68±11.24 | 24.94±19.70 | 20.05±14.14 | 77.00±42.48 | 115.10±35.61 | 0.63±0.07 | 0.80±0.04 |
| | Min | 3 | 4 | 5 | 3 | 4 | 16 | 56 | 0.52 | 0.73 |
| | Max | 34 | 50 | 39 | 59 | 47 | 183 | 185 | 0. 80 | 0.88 |

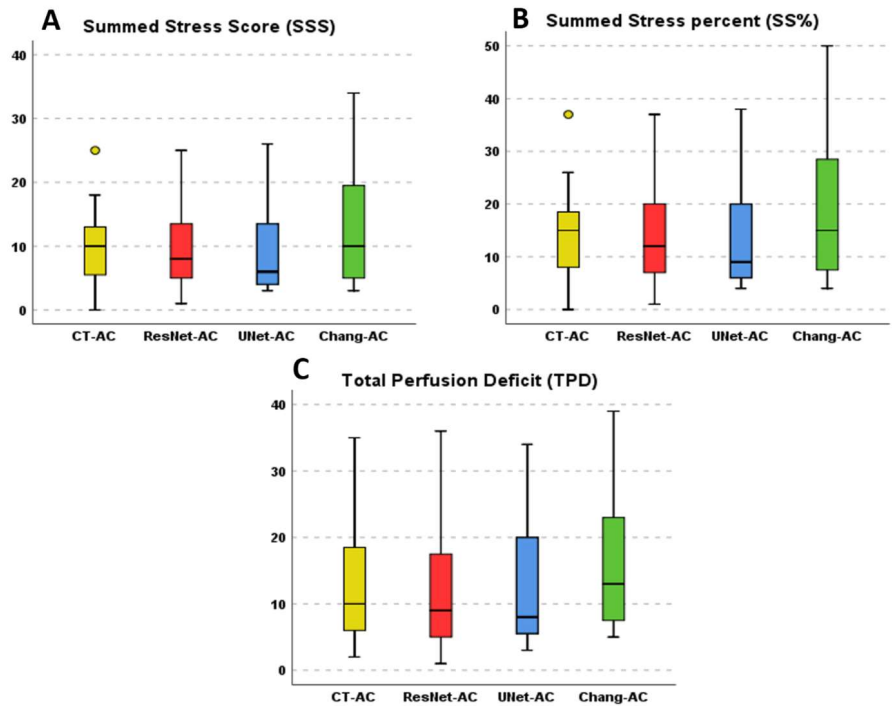

**Figure 7.** quantitative assessment of perfusion in the MPI-SPECT images; including A) summed stress score (SSS), B) summed stress percent (SS%), and C) total perfusion deficit (TPD%) for the SPECT CT-AC, SPECT ResNet-AC, SPECT UNet-AC, and SPECT Chang-AC images.

## 4. DISCUSSION

In this work, we have investigated the possibility of direct attenuation correction of SPECT myocardial perfusion imaging with $^{99m}$Tc-sestamibi in the image domain using deep learning-based models. The quantitative results obtained from the 19 patients in the external dataset demonstrated that the deep learning-based methods (namely ResNet and UNet) were able to produce attenuation corrected myocardial perfusion SPECT images featuring an excellent agreement with the reference CT-based attenuation corrected MPI-SPECT images. The Change method was also implemented to provide a baseline for an insightful performance assessment of the deep learning solutions. Though the Chang method is less applicable to the thorax and pelvis regions due to their high heterogeneity attenuating medium, this method would enable to partly recover the signal loss caused by the attenuated or scattered photons. As such, this method would relatively provide improved image quality/contrast (compared to non-AC image) to be considered as an alternative approach for the task of attenuation correction in cases the additional radiation dose of CT is not acceptable [43]. The deep learning models remarkably outperformed the Chang AC methods considering the quantitative metrics as well as the clinical assessment.

The SPECT/CT hybrid systems which routinely benefit from CT-based attenuation correction in clinical practice, enable quantitative MPI-SPECT imaging owning to the accurate, patient-specific, and low noise AC map. However, this hybrid imaging modality suffers from some limitations such as additional patient radiation exposure for CT scan and the misregistration between CT and SPECT images that may cause quantitative uncertainty or image artifact [17, 44]. The deep learning-based attenuation correction in the image domain would be able to eliminate the need for any transmission scanning or CT-derived AC maps. Deep learning-based AC in the image domain evaluated for PET imaging has shown a promising capability to account for the misalignment between emission and transmission data due to the respiratory or patient bulk motions [35, 45].

In recent years, deep learning methods have been increasingly applied for attenuation correction in PET images and presented promising results for the generation of AC maps from MR images or direct AC of non-attenuation corrected PET in the image domain [31, 33, 46]. However, in SPECT imaging fewer studies have been conducted on the deep learning-based AC due to less application of the SPECT/MR hybrid systems in clinical practice [20, 29]. In this regard, Shi *et al.* proposed a novel framework wherein both photopeak and scatter energy windows of an MPI-SPECT scan are reconstructed to be fed into a generative adversarial network (GAN). Then, the GAN model would be able to estimate patient-specific AC map from scatter and photopeak images relying on the existing complementary information in these images. The quantitative analysis of this approach performed on 25 patients revealed a normalized mean absolute error (NMAE) of 3.60% ± 0.85% between the synthetic and CT-based attenuation maps leading to an accurate attenuation correction for myocardial perfusion SPECT images [19]. Nguyen *et al.* in a recent study, proposed a 3DUnet-Gan network (the 3D GAN network model which uses 3D UNet as the generator), generating attenuation corrected MPI-SPECT from non-attenuation corrected SPECT as input. The results of this work showed that their proposed model could generate synthetic SPECT-AC images with SSIM = 0.945% and NMAE = 0.034 compared to the reference SPECT CT-AC images [47].

The levels of errors observed in this study are in agreement with the previous studies which indicate that the deep learning solutions would able to achieve the synthesis of MPI-SPECT $^{99m}$Tc-sestamibi images with clinically tolerable errors. Such a deep learning AC framework could be employed in SPECT-only or SPECT/MR scanner to achieve reliable quantitative imaging. Nevertheless, further studies are required to evaluate the feasibility of this method on different radiotracers, regions of the body, acquisition protocols, and a broad range of defects/diseases.

## 5. CONCLUSION

This study investigated the direct attenuation correction of the MPI-SPECT $^{99m}$Tc-sestamibi images in the image domain using the deep learning-based algorithms. The quantitative metrics and clinical evaluation revealed an excellent agreement of the deep learning-based synthetic SPECT images with the reference SPECT CT-AC images with clinically tolerable errors.